\documentclass[pra,amsmath,amssymb]{revtex4}  %superscriptaddress
\usepackage{graphics}
\usepackage{epsfig}
\usepackage{bbm}
\usepackage{amssymb}
\usepackage[latin1]{inputen c} % Umlaute!

\newcommand{\be}{\begin{equation}}
\newcommand{\ee}{\end{equation}}
\newcommand{\eea}{\end{eqnarray}}
\newcommand{\bea}{\begin{eqnarray}}

\newcommand{\exs}[1]{\ensuremath{\langle{#1}\rangle}}

\newcommand{\WW}{\ensuremath{\mathcal{W}}}

\newcommand{\ketbra}[1]{\ensuremath{| #1 \rangle \langle #1 |}}

\newcommand{\ket}[1]{\ensuremath{|#1\rangle}}

\newcommand{\kommentar}[1]{}
\newcommand{\trace}{{\rm Tr}}

\newcommand{\EQ}[1]{(\ref{#1})}
\newcommand{\EQEQ}[1]{Equation~(\ref{#1})}
\newcommand{\REF}{}

\newcommand{\br}{\hline}
\newcommand{\fl}{}
\newcommand{\ack}{\section*{Acknowledgments}}

\begin{document}
\title[Practical methods for
witnessing genuine multi-qubit entanglement...]{Practical methods for
witnessing genuine multi-qubit entanglement in the vicinity of symmetric states}

\date{\today}

\begin{abstract}
We present general numerical methods to construct witness operators for entanglement detection and estimation of the fidelity. Our methods are applied to detecting entanglement in the vicinity of a six-qubit Dicke state with three excitations and also to further entangled symmetric states. All our witnesses
are designed to keep the measurement effort small. We present also general results on the efficient local decomposition of permutationally invariant operators, which makes it possible to measure projectors to symmetric states efficiently.
\end{abstract}

\author{G\'eza T\'oth$^{1,2,3}$,Witlef Wieczorek$^{4,5}$, Roland Krischek$^{4,5}$, Nikolai Kiesel$^{4,5,6}$, Patrick Michelberger$^{4,5}$, Harald Weinfurter$^{4,5}$}

\affiliation{$^1$ Department of Theoretical Physics, The University of the Basque Country,
P.O. Box 644, E-48080 Bilbao, Spain }
\affiliation{$^2$ IKERBASQUE, Basque Foundation for
Science, E-48011 Bilbao, Spain}
\affiliation{$^3$ Research Institute for Solid State Physics and Optics,
Hungarian Academy of Sciences, P.O. Box 49, H-1525 Budapest,
Hungary}
\affiliation{$^4$ Max-Planck-Institut f\"ur Quantenoptik, Hans-Kopfermann-Strasse 1, D-85748 Garching, Germany}
\affiliation{$^5$ Fakult\"at f\"ur Physik, Ludwig-Maximilians-Universit\"at, D-80799 Garching, Germany}
\affiliation{$^6$ Institute for Quantum Optics and Quantum Information (IQOQI), Austrian
Academy of Sciences, Boltzmanngasse 3, A-1090 Vienna, Austria}

\email{toth@alumni.nd.edu}

%\submitto{\NJP}

\maketitle

% 03.65.Ud                  Entanglement and quantum nonlocality
% 03.67.Mn                  Entanglement in Quantum Information
% 03.67.-a                  Quantum information
% 03.67.Lx                  Quantum computation
% 02.70.-c                  Computational techniques
% 05.50.+q                  Lattice theory and statistics (Ising, Potts, etc.)
%                           (see also 64.60.Cn Order-disorder transformations,
%                           and 75.10.Hk Classical spin models)
% 42.50.Dv                  Quantum state engineering and measurements
%                           (see also 03.65.Ud Entanglement and quantum
%                           nonlocality, e.g., EPR paradox, Bells inequalities, GHZ states, etc.)

%%%%%%%%%%%%%%%%%%%%%%%%%%%%%%%%%%%%%%%%%%%%%%%%%%%%%%%%%%%%%%%%%%%%%%

\section{Introduction}

Entanglement plays a central role in quantum mechanics and in quantum information processing applications \cite{reviews}. Moreover, it appears also as the main goal in today's quantum physics experiments aiming to create various quantum states \cite{GT08}. For example, entanglement has been realized with photonic systems using parametric down-conversion and conditional detection \cite{PB00,BE04,KS05,KS07,WS08,LZ07,GL08}, with trapped cold ions \cite{SK00,LK05,HH07}, in cold atomic ensembles \cite{KB98}, in cold atoms in optical lattices \cite{MG03} and in diamond between the electron and nuclear spins \cite{NM08}. These experiments
aimed at creating entangled states. Entanglement makes it possible for some quantum algorithms (e.g., prime factoring, search in a database) to outperform their classical counterparts. Entangled
particles are needed for quantum teleportation and other quantum communication protocols.
Moreover, the creation of large entangled states might lead to new insights about how a classical macro-world emerges from a quantum micro-world.

In a multi-qubit experiment, typically the full density matrix is not known, and only few measurements can be made, yet one would still like to ensure that the prepared state is entangled. One possibility is applying entanglement witnesses \cite{T02,LK00}. These are observables that have a positive expectation value for separable states, while for some entangled states their expectation value is negative. Since these witness operators are multi-qubit operators, they typically cannot be measured directly and must be decomposed into the sum of locally measurable operators, which are just products of single-qubit operators \cite{GH02,GH03,BE04}.

For many quantum states, like the Greenberger-Horne-Zeilinger (GHZ, \cite{GH90}) states and the cluster states \cite{BR01} such a decomposition of projector-based witness operators seems to be very difficult: The number of terms in a decomposition to a sum of products of Pauli matrices increases rapidly with the number of qubits. However, practically useful entanglement witnesses with two measurement settings can be constructed for such states \cite{KS05,TG05}. It also turned out that there are decompositions of the projector for GHZ and W-states in which the increase with the number of qubits is linear \cite{toolbox}.

However, optimal decomposition of an operator is a very difficult, unsolved problem. Moreover, in general,
it is still a difficult task to construct efficient entanglement witnesses for
a given quantum state. For that, typically we need to obtain the maximum of some operators for product states. In most of the cases we would like to detect genuine multipartite entanglement. For that, we need to obtain the maximum of these operators for biseparable states, which is again a very hard problem.

In this paper our goal is to design witnesses that make it possible to detect genuine multipartite entanglement with few measurements, and also to estimate the fidelity of an experimentally prepared state with respect to the target state. Here three strategies are applied to find an experimentally realizable witness. (i) The first strategy is based on measuring the projector-based witness
\be
\WW^{(P)}=\textrm{const.}\cdot\openone-\ketbra{\Psi}
\ee
for the detection of genuine multipartite entanglement. $\ket{\Psi}$ is the target state of the experiment. For reducing experimental effort, the aim is to find an efficient decomposition of the projector.
(ii) The second strategy is to find a witness that needs much fewer measurements than the projector witness, but the price for that might be a lower robustness against noise.
The search for such a witness can be simplified if we look for a witness $\WW$ such that
\be
\WW-\alpha \WW^{(P)}\ge0\label{WWWP}
\ee
for some $\alpha>0.$ Such a witness is guaranteed to detect genuine multi-qubit entanglement.
The advantage of this approach is that the expectation value of $\WW$ can be used to find a lower bound on the fidelity.
(iii) The third strategy is to find a witness independent from the projector witness. In this case one has to find an easily measurable operator whose expectation value takes its maximum for the target state. Then, one has to find the maximum of this operator for biseparable states. Any state that has an operator expectation value larger than that is genuine multipartite entangled.

For the optimization of entanglement witnesses for small experimental effort and large robustness to noise, we use semidefinite programming \cite{DP02,BV04,EH04,JN07,HE08,WP09}. Our methods can efficiently be used for multi-qubit systems with up to about $10$ qubits. This is important, since there are many situations where semidefinite programming could help theoretically, but in practice the calculations cannot be carried out even for systems of modest size.

We use our methods to design witnesses detecting entanglement in the vicinity of symmetric Dicke states.
An $N$-qubit symmetric Dicke state with $m$ excitations is
defined as \cite{Dicke,SG03}
\begin{equation}
\ket{D_N^{(m)}}:=\bigg(\begin{array}{c}N \\
m\end{array}\bigg)^{-\frac{1}{2}}\sum_k \mathcal{P}_k
(\ket{1_1,1_2,...,1_m,0_{m+1},...,0_N}), \label{sd}
\end{equation}
where $\sum_k \mathcal{P}_k(.)$ denotes summation over all distinct permutations of the
spins. $\ket{D_N^{(1)}}$ is the well known $N$-qubit W state.
The witnesses we will introduce in the following have already been used in the photonic experiment described in \REF\cite{dicke6experiment},
aiming to observe a $\ket{D_6^{(3)}}$ state \cite{dicke6experiment,dicke6experiment2}.
We show that genuine multi-qubit entanglement can be detected and the fidelity with respect to the above highly entangled state can efficiently be estimated with two and three measurement settings, respectively.
 As a byproduct, we will also derive an upper bound for the number of settings needed to measure any permutationally invariant operator. We show that such operators can efficiently be measured even for large systems.

The structure of our paper is as follows. In section~2, we present the basic methods for constructing witnesses.
In section~3, we use these methods for constructing witnesses to detect entanglement in the vicinity of a  six-qubit symmetric Dicke state with three excitations. In section~4, we present witnesses for states obtained from the above state by measuring some of the qubits. In Appendix~A, we summarize the tasks that can be solved by semidefinite programming, when looking for suitable entanglement witnesses. In Appendix~B, we summarize some of the relevant numerical routines of the QUBIT4MATLAB 3.0 program package \cite{QUBIT4MATLAB}. In Appendix~C, we present entanglement conditions for systems with $5-10$ qubits that will be relevant in future experiments.

\section{Basic definitions and general methods}

A multi-qubit quantum state is entangled if it cannot be written as a convex combination of product states.
However, in a multi-qubit experiment we would like to detect genuine multi-qubit entanglement \cite{AB01}:
The presence of such entanglement indicates that all the qubits are entangled with each other, not only some of them.
We will now need the following definitions:\\
{\bf Definition 1. } A pure multi-qubit quantum state is called {\bf biseparable} if it can be
written as the tensor product of two, possibly entangled, multi-qubit states
\be
\ket{\Psi}=\ket{\Psi_1}\otimes\ket{\Psi_2}.
\ee
%Here $\ket{\Psi}$ is an $N$-qubit state, while $\ket{\Psi_1}$ and $\ket{\Psi_2}$ are $N_1$-qubit and
%$N_2$-qubit
%states, respectively, such that $N=N_1+N_2.$
A mixed state is called biseparable, if it can be obtained by mixing pure biseparable states.
If a state is not biseparable then it is called {\bf genuine multi-partite entangled}. In this paper we will consider witness operators that detect genuine multipartite entanglement.

{\bf Definition 2. } While an entanglement witness is an observable, typically it cannot be measured directly. This is because in most experiments only local measurements are possible. At each qubit $k$ we are able to measure a single-qubit operator $M_k,$ which we can do simultaneously at all the qubits. If we repeat such measurements, then we obtain the expectation values of $2^N-1$ multi-qubit operators. For example, for $N=3$ these are $M_1\otimes\openone\otimes\openone,\openone\otimes M_2\otimes\openone,\openone\otimes\openone\otimes M_3,M_1\otimes M_2\otimes\openone, M_1\otimes\openone\otimes M_3,\openone\otimes M_2\otimes M_3,M_1\otimes M_2\otimes M_3.$ The set of single-qubit operators measured is called {\bf measurement setting} \cite{BE04} and it can be given  as $\{M_1,M_2,M_3,...,M_N\}.$ When we consider an entanglement condition, it is important to know, how many measurement settings are needed for its evaluation.

{\bf Definition 3. } Many experiments aim at preparing some, typically pure quantum state $\varrho.$
An entanglement witness is then designed to detect the entanglement of this state. However, in real experiments such a state is never produced perfectly, and the realized state is mixed with noise as
given by the following formula
\be
\varrho_{\rm noisy}(p_{\rm noise})=(1-p_{\rm noise})\varrho+p_{\rm noise}\varrho_{\rm noise}, \label{noise}
\ee
where $p_{\rm noise}$ is the ratio of noise and $\varrho_{\rm noise}$ is the noise.
If we consider white noise then $\varrho_{\rm noise}=\openone/2^N.$
The  {\bf noise tolerance} of a witness $\WW$ is characterized by the largest  $p_{\rm noise}$ for which we still have $\trace(\WW\varrho_{\rm noisy})<0.$

In this paper, we will consider three possibilities for detecting genuine multi-qubit entanglement, explained in the following subsections. Later, we will use these ideas to construct various entanglement witnesses.

\subsection{Projector witness}

A witness detecting genuine multi-qubit entanglement in the vicinity of a pure state $\ket{\Psi}$ can be constructed with the projector as
\begin{equation}
\WW_{\Psi}^{(P)}:=\lambda_\Psi^2\openone-\ketbra{\Psi},\label{overlap}
\end{equation}
where $\lambda$ is the maximum of the Schmidt coefficients for $\ket{\Psi}$, when all bipartitions are considered \cite{BE04}.
%In other words, $\lambda$ can be obtained from $\Psi$ through simple calculations.
For the states considered in this paper, projector-based witnesses are given by \cite{T07,BE04,HH07}
\begin{eqnarray}
\WW_{\rm D(N,N/2)}^{(P)}&:=&\tfrac{1}{2}\tfrac{N}{N-1}\openone-\ketbra{D_N^{(N/2)}}, \label{WPD}\\
\WW_{\rm D(N,1)}^{(P)}&:=&\tfrac{N-1}{N}\openone-\ketbra{D_N^{(1)}}. \label{WW}
\end{eqnarray}

These witnesses must be decomposed into the sum of locally measurable terms.
For this decomposition, the following observations will turn out to be very important.\\
{\bf Observation 1.} A permutationally invariant operator $A$ can always be
decomposed as \cite{tensor_rank}
 \be
   A=\sum_n c_n a_n^{\otimes N}, \label{dec}
 \ee
where $a_n$ are single qubit operators, and such a decomposition can straightforwardly be obtained. \\
{\it Proof.} Any permutationally invariant multi-qubit operator $A$
can be decomposed as
 \be
   A=\sum_n c_n \sum_k P_k ( B_{n,1} \otimes B_{n,2} \otimes B_{n,3} \otimes ...\otimes B_{n,N} ) P_k,\label{dec2}
 \ee
where $B_{n,m}$ are single qubit operators, $c_n$ are constants, and $P_k$ are
the full set of operators permuting the qubits. For odd $N,$ we can use the identity
\begin{eqnarray}
   \fl&&\sum_k P_k ( B_{n,1} \otimes B_{n,2} \otimes B_{n,3} \otimes ...\otimes B_{n,N} ) P_k\nonumber\\
      &&\;\;\;\;\;\;\;\;\;\;\;\;\;\;\;\;\;\;\;\;\;=2^{-(N-1)} \sum_{
   \begin{array}{c}
   s_1,s_2,...=\pm 1,\\
   s_1s_2s_3\cdot\cdot\cdot s_N=+1
   \end{array}
   } ( s_1B_{n,1} +s_2 B_{n,2} +s_3 B_{n,3}+ ...)^{\otimes N}.\nonumber\\\label{dec3}
 \end{eqnarray}
 Substituting \EQ{dec3} into \EQ{dec2}, we obtain a decomposition of the form \EQ{dec}.
 \EQEQ{dec3} can be proved by carrying out the summation and expanding the brackets.
 Due to the  $s_1s_2s_3\cdot\cdot\cdot s_N=+1$ condition, the coefficient of  $B_{n,1} \otimes B_{n,2} \otimes B_{n,3} \otimes ...\otimes B_{n,N}$ is $1.$ The coefficient of terms like $B_{n,1} \otimes B_{n,1} \otimes B_{n,3} \otimes ...\otimes B_{n,N},$ that is, terms containing one of the variables more than once
 is zero. For even $N,$ a similar proof can be carried out using
 \footnote{A similar decomposition with continuous number of terms is of the form
$\sum_k P_k (B_1 \otimes B_2 \otimes ...)P_k \propto \int_{ \phi_k\in[0,2\pi]}[e^{i\phi_1}B_1+e^{i\phi_2}B_2+...+e^{i\phi_{N-1}}B_{N-1}+e^{-i(\phi_1+\phi_2+...+\phi_{N-1})}B_N]^{\otimes N}d\phi_1 d\phi_2...d\phi_{N-1}.$ Such a construction has been used for the $N=2$ case in Keilmann T and Garc\'{\i}a-Ripoll J J 2008 \emph{Phys. Rev. Lett.} {\bf 100} 110406.}
 \begin{eqnarray}
   \fl&&\sum_k P_k ( B_{n,1} \otimes B_{n,2} \otimes B_{n,3} \otimes ...\otimes B_{n,N} ) P_k\nonumber\\
   &&\;\;\;\;\;\;\;\;\;\;\;\;\;\;\;\;\;\;\;\;\;=2^{-(N-1)} \sum_{
   \begin{array}{c}
   s_1,s_2,...=\pm 1,\\
   s_1s_2s_3\cdot\cdot\cdot s_N=+1
   \end{array}
   } s_1( B_{n,1} +s_2 B_{n,2} +s_3 B_{n,3}+ ...)^{\otimes N}.\nonumber\\\label{dec3b}
 \end{eqnarray}
Next, we give two examples for the application of \EQ{dec3} and \EQ{dec3b} for the decomposition of simple expressions \bea
\sum_k P_k (\sigma_x\otimes\sigma_y)P_k&=&\tfrac{1}{2}\bigg\{(\sigma_x+\sigma_y)^{\otimes 2}-(\sigma_x-\sigma_y)^{\otimes 2}\bigg\},\\
\sum_k P_k (\sigma_x\otimes\sigma_y\otimes\sigma_z)P_k &=& \tfrac{1}{4}\bigg\{(\sigma_x+\sigma_y+\sigma_z)^{\otimes 3}+(\sigma_x-\sigma_y-\sigma_z)^{\otimes 3}\nonumber\\&+&(-\sigma_x-\sigma_y+\sigma_z)^{\otimes 3}+(-\sigma_x+\sigma_y-\sigma_z)^{\otimes 3}\bigg\},
\eea
where $\sigma_k$ are the Pauli spin matrices. While the first example does not reduce the number of settings needed, the second example reduces the number of settings from $6$ to $4.$

Next, we present a method to get efficient decompositions
for permutationally invariant operators.\\
{\bf Observation 2.} Any $N$-qubit permutationally invariant  operator $A$ can be
measured with at most \be \mathcal{L}_N=\tfrac{2}{3}N^3+N^2+\tfrac{4}{3}N
\label{LN}\ee local measurement settings, using \EQ{dec3} and \EQ{dec3b}. \\
{\it Proof.} We have to decompose first $A$ into the sum
of Pauli group elements as
\be
A=\sum_{i,j,m:\;i+j+m\le N} c_{ijm} \sum_k P_k ( \sigma_x^{\otimes i}\otimes\sigma_y^{\otimes j}\otimes\sigma_z^{\otimes m}\otimes\openone^{\otimes (N-i-j-m)}) P_k,
\ee
where $c_{ijm}$ are some constants. Then, such a decomposition can be transformed
into another one of the form \EQ{dec}, using \EQ{dec3} and \EQ{dec3b}.
All of the settings needed
are of the form $\{a,a,a,...,a\}$ where $a=n_x\sigma_x+n_y\sigma_y+n_z\sigma_z,$
$n_k$ are integer and $1 \le \sum_k \vert n_k \vert \le N.$
Simple counting leads to an upper bound $\mathcal{L}_N$ for the number of settings given in  \EQ{LN}.
Here we considered that $(n_x,n_y,n_z)$ and $(-n_x,-n_y,-n_z)$ describe the same setting.
An even better bound can be obtained using that $(n_x,n_y,n_z)$ and $(cn_x,cn_y,cn_z)$ for some $c\ne 0$
represent the same setting. An algorithm based on this leads to the bounds
$\mathcal{L}_N'=9,25,49, 97, 145, 241, 337, 481, 625$ for
$N=2,3,...,10$ qubits, respectively.

For the projector $\ketbra{D_N^{(N/2)}},$ the decomposition to Pauli group elements contain
only terms in which each Pauli matrix appears an even number of times.
Hence, all of the settings needed are of the form
$\{a,a,a,...,a\}$ where $a=2n_x\sigma_x+2n_y\sigma_y+2n_z\sigma_z,$
$n_k$ are integer and $1 \le \sum_k \vert n_k \vert \le N/2.$ For this reason, $\mathcal{L}_{N/2}$ and $\mathcal{L}_{N/2}'$ are
upper bounds for the number of settings needed to measure this operator.

Let us discuss the consequences of Observations 1 and 2. They essentially state that the number of settings needed to measure a permutationally invariant operator scales only polynomially with the number of qubits.
This is important since for operators that are not permutationally invariant, the scaling is known to be exponential \cite{phdguhne}. Moreover, even if we can measure only correlation terms of the form $a^{\otimes N},$ we can measure any permutationally invariant operator.

\subsection{Witnesses based on the projector witness}
\label{sec_projbased}

We can construct witnesses that are easier to measure than the projector witness,
but they are still based on the projector witness.
We use the idea mentioned in the introduction.
If $\WW^{(P)}$ is the projector witness and \EQ{WWWP}
is fulfilled
for some $\alpha>0,$ then $\WW$ is also a witness.
This is because $\WW$ has a negative expectation value only for states for which
$\WW^{(P)}$ also has a negative expectation value.
The advantage of obtaining witnesses this way is that we can have a lower bound on the fidelity
from the expectation value of the witness as
\be
\trace(\varrho\ketbra{\Psi})\ge\lambda_\Psi^2-\tfrac{1}{\alpha}\trace(\WW\varrho).\label{fid_est}
\ee

We will look for such witnesses numerically,
such that the noise tolerance of the witness be the largest possible. This search can be simplified by the following observation.\\
{\bf Observation 3.} Since we would like to construct a witness detecting genuine multiqubit
entanglement
in the vicinity of a permutationally invariant state, it is enough to consider witness
operators that are also permutationally invariant. \\
{\it Proof.} Let us consider a witness operator that detects entanglement in the vicinity of
a permutationally invariant state $\varrho$ and its expectation value takes its minimum for $\varrho.$
Then, based on \EQ{noise}, the witness $\WW$ detects entanglement if
\be
p_{\rm noise}>\frac{{\rm Tr}(W\varrho)} {{\rm Tr}(W\varrho)-{\rm Tr}(W\varrho_{\rm noise})}.\label{robustness}
\ee
For a permutationally invariant state $\varrho,$ we have
$
\varrho=\tfrac{1}{N_P}\sum_k P_k \varrho P_k,
$
where $N_P$ is the number of different permutation operators $P_k$.
We assume that the same holds also for $\varrho_{\rm noise}.$
Let us define the permutationally invariant operator $\WW'=\tfrac{1}{N_P}\sum_k P_k \WW P_k.$
The operator $\WW'$ is non-negative on all biseparable states since
\begin{equation}
\inf_{\varrho\in \mathcal{B}} \trace(W\varrho)=\tfrac{1}{N_P}\sum_k \inf_{\varrho\in \mathcal{B}} \trace(\WW P_k \varrho P_k)\le
\inf_{\varrho\in \mathcal{B}} \trace(\WW' \varrho),
\end{equation}
where $\mathcal{B}$ is the set of biseparable states.
Hence, $\WW'$ is a witness detecting genuine multipartite entanglement.
Since we have $\trace(\WW\varrho)=\trace(\WW'\varrho),$
and $\trace(\WW\varrho_{\rm noise})=\trace(\WW'\varrho_{\rm noise}),$
the robustness to noise of $\WW'$ is identical to that of $\WW.$
Hence, it is sufficient to look for witnesses
that are permutationally invariant.

We will first consider measuring the $\{\sigma_x,\sigma_x,\sigma_x,\sigma_x,\sigma_x,\sigma_x\}$ and $\{\sigma_y,\sigma_y,\sigma_y,\sigma_y,\sigma_y,\sigma_y\}$ settings, where $\sigma_l$ are the Pauli spin matrices.
This we call the two-setting case.
Then we will consider measuring also the $\{\sigma_z,\sigma_z,\sigma_z,\sigma_z,\sigma_z,\sigma_z\}$ setting, which we call the three-setting case. Due to Observation 3,
we consider only permutationally invariant witnesses.
Such witnesses can be written as
\be
  \WW(\alpha_0,\{\alpha_{ln}\}):=\alpha_0\cdot\openone+\sum_{l=x,y,z}\sum_{n=1}^N \alpha_{ln} \sum_k \mathcal{P}_k [\sigma_l^{\otimes n}\otimes\openone^{\otimes (N-n)}], \label{permform}
\ee
where the summation is over all distinct permutations, and $\alpha_0$ and $\alpha_{ln}$ are some constants. We will consider a simpler but equivalent formulation
\be
  \WW(c_0,\{c_{ln}\}):=c_0\cdot\openone+\sum_{l=x,y,z}\sum_{n=1}^N c_{ln} J_l^n, \label{witJ}
\ee
where $c_0,$ $c_{ln}$ are the coefficients of the linear combination defining the witness and
$J_l$ are the components of the total angular momentum given as
 \be
 J_l=\frac{1}{2} \sum_{k=1}^N \sigma_l^{(k)}.
 \ee
Here $\sigma_l^{(k)}$ denotes a Pauli spin matrix acting on qubit $(k).$

Finally, if we consider detecting entanglement in the vicinity of $\ket{D_N^{(N/2)}}$ states, then further simplifications can be made. For this state and also
for the completely mixed state all odd moments of $J_l$
have a zero expectation value. For any witness of the form \EQ{witJ}, the maximum for biseparable states does not change if we flip the sign of $c_{ln}$ for all odd $n.$ Hence, following from an argument similar to the one in Observation 3 concerning permutational symmetry, it is enough to consider only even powers of $J_l$ in our witnesses.

\subsection{Witnesses independent from the projector witness.}
\label{sec_w_indep_pw}

In general, we can also design witnesses without any relation to the projector witness.
We can use an easily measurable operator $M$ to make a witness of the form
\begin{equation}
\WW:=c\openone-M, \label{WWcM}
\end{equation}
where $c$ is some constant.
To make sure that
\EQ{WWcM} is a witness for genuine multipartite entanglement, i.e,
$\exs{\WW}$ is positive on all biseparable states,
we have to set $c$ to
\begin{equation}
c=\max_{\ket{\Psi} \in \mathcal{B}} \exs{M}_{\ket{\Psi}}, \label{optbisep}
\end{equation}
where $\mathcal{B}$ is the set of biseparable states.
The optimization needed for \EQ{optbisep} can be done analytically. For example, for the $\ket{D_{4}^{(2)}}$ state a witness has been presented that detects genuine four-qubit entanglement by measuring second moments of angular momentum operators \cite{T07}. However, analytical calculations become exceedingly difficult as the number of qubits increases.

The optimization can also be done numerically, but
one cannot be sure that simple numerical optimization
finds the global maximum. (See Appendix B for a reference to such a MATLAB program.)
Semidefinite programming is known to find the global optimum, but
the optimization task \EQ{optbisep} cannot be solved directly by semidefinite programming.
Instead of looking for the maximum for biseparable states, using semidefinite programming, we can look for the maximum for states that have a positive partial transpose \cite{ppt,DP02}.
(See Appendices A and B.)
This way we can obtain
\begin{equation}
c':=\max_I \max_{\varrho\ge 0, \varrho^{T_I}\ge 0} \exs{M}_{\varrho}, \label{optbisep2}
\end{equation}
for which $c'\ge c.$ The first maximization is over all bipartitions $I.$ Thus, when putting $c'$ into the place of $c$ in \EQ{WWcM}, we obtain a witness that detects only genuine multipartite entanglement. In many cases simple numerics show that $c=c'.$ In this case our witnesses are optimal in the sense that some biseparable state gives a zero expectation value for
these witnesses.

Finally, let us discuss how to find the operator $M$ in \EQ{WWcM} for a two- or a three-setting witness, in particular, for detecting entanglement in the vicinity of $\ket{D_N^{(N/2)}}.$ Based on section~\ref{sec_projbased},
we have to look for an operator that contains only even powers of $J_l.$
Hence, the general form of a two-setting witness with moments up to second order is
\be
   \WW_{\rm D(N,N/2)}^{(I2)}:=c_{DN}-(J_x^2+J_y^2), \label{WI2}
\ee
where $c_{DN}$ is a constant \cite{structure}.
The coefficients of $J_x^2$ and $J_y^2$ could still be
 different, however, this would not lead to witnesses with a better robustness to noise.

For other symmetric Dicke states, based on similar arguments, a general form of a witness
containing moments of $J_l$ up to second order  such that it takes its minimum for
$\ket{D_N^{(m)}}$ is
of the form
\be
   \WW_{\rm D(N,m)}^{(I3)}:=c_q-(J_x^2+J_y^2)+q(J_z-\exs{J_z}_{\ket{D_N^{(m)}}})^2, \label{WI3}
\ee
where $c_q$ and $q$ are constants.
For the witnesses described in this section, the optimization process is more time-consuming
than for the witnesses related to the projector witness. Because of that we presented witnesses
of the above type that are constructed only with the first and second moments of
the angular momentum operators, and thus contain few free parameters.

\section{Witnesses for a six-qubit Dicke state with three excitations}

In this section, we will consider entanglement detection close to a six-qubit symmetric Dicke state with three excitations, denoted as $\ket{D_6^{(3)}}.$ There are several proposals for creating Dicke states
in various physical systems \cite{UF03,DK03,TZ07,SH04}.

\subsection{Witnesses based on the projector witness}

\subsubsection{Two-setting witness}
\label{sec_ts_bp}

Let us consider the two-setting case
and define first the optimization problem we want to solve.
We would like to look for the witness $\WW$ with
the largest noise tolerance that fulfills the following
requirements:
\begin{itemize}
\item[(i)] $\WW$ is a linear combination of certain basis operators $B_k,$
that is,  $\WW=\sum_k c_k B_k,$
\item[(ii)]  $\WW-\alpha \WW_{\rm D(6,3)}^{(P)}\ge 0$ with some $\alpha>0.$
\end{itemize}
For the two-setting case we set $\{B_k\}=\{\openone,
J_x^2,J_y^2,J_x^4,J_y^4,J_x^6,J_y^6\}.$
The second condition makes sure that $\WW$ is also a witness detecting genuine multipartite entanglement.

Note that {\it any} optimization algorithm can be used for looking for $\WW.$
Even if we do not find the global optimum, that is,
the witness with the largest possible robustness to white noise,
$\WW$ is still a witness detecting genuine multipartite entanglement. However, semidefinte programming can be used
to find the global optimum. (See Appendix A.)
The two-setting witness obtained this way is %\cite{alternative_form}
 \be
 \WW_{\rm D(6,3)}^{(P2)}:=7.75\cdot\openone -\tfrac{35}{18}(J_x^2+J_y^2)+\tfrac{55}{72}(J_x^4+J_y^4)-\tfrac{5}{72}(J_x^6+J_y^6),
 \ee
which tolerates white noise if $p_{\rm noise}<0.1391.$ Straightforward calculation shows that
$\WW_{\rm D(6,3)}^{(P2)}-2.5\WW^{(P)}\ge 0.$ Based on \EQ{fid_est}, $(0.6-{\exs{\WW_{\rm D(6,3)}^{(P2)}}}/{2.5})$ bounds the fidelity from below.

\subsubsection{Three-setting witness}

Similarly we can look for the optimal witness for the three-setting case.
The result is %\cite{alternative_form}
 \begin{eqnarray}
 \WW_{\rm D(6,3)}^{(P3)}&:=&1.5\cdot\openone -\tfrac{1}{45}(J_x^2+J_y^2)+\tfrac{1}{36}(J_x^4+J_y^4)-\tfrac{1}{180}(J_x^6+J_y^6)\nonumber\\
 &+&\tfrac{1007}{360}J_z^2-\tfrac{31}{36}J_z^4+\tfrac{23}{360}J_z^6.\label{WD63P3}
 \end{eqnarray}
White noise is tolerated if $p_{\rm noise}<0.2735.$
It is easy to check that $\WW$ is a witness as
$\WW_{\rm D(6,3)}^{(P3)}-2.5\WW^{(P)}\ge 0.$

\begin{figure}
\centerline{ \epsfxsize 9cm \epsffile{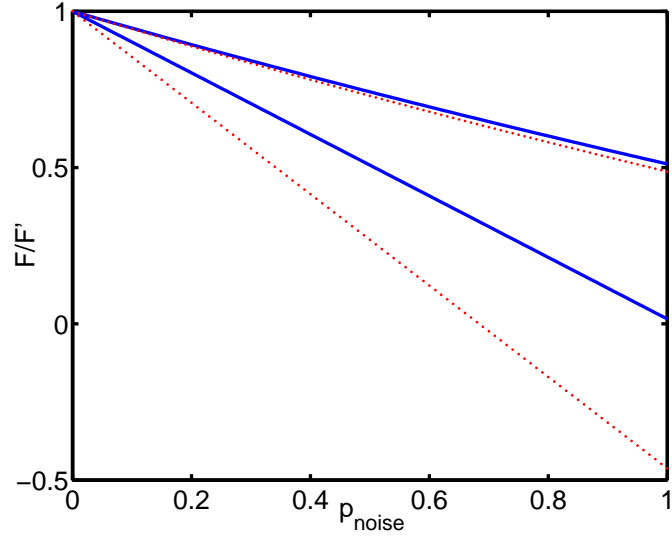}}
\caption{The fidelity $F$ vs. noise (solid) and the fidelity estimate $F'$ vs. noise (dotted), for the white noise (bottom two curves) and for the non-white noise \EQ{nw} (top two curves). For the fidelity estimate, the three-setting witness \EQ{WD63P3} was used.} \label{fig_d6_noise}
\end{figure}

Based on \EQ{fid_est}, the expectation value of this witness can be used to bound the
fidelity as $F \ge 0.6-{\exs{\WW_{\rm D(6,3)}^{(P3)}}}/{2.5}=:F'.$
Here we will demonstrate how well the fidelity estimation works for our witness for noisy states.
We consider first white noise, then non-white noise of the form
 \be
 \varrho_{\rm noisy}^{\rm (NW)}:=p_{D63}\ketbra{D_6^{(3)}}+\tfrac{1-p_{D63}}{2}\left(\ketbra{D_6^{(2)}}+\ketbra{D_6^{(4)}}\right), \label{nw}
 \ee
with $p_{D63}=4/7,$ which is one of the relevant types of noise for the experiment of
\REF\cite{dicke6experiment}. Note that the noise contains the original state $\ketbra{D_6^{(3)}}.$ The results are shown in figure~\ref{fig_d6_noise}. For the non-white noise \EQ{nw}, the fidelity estimate based on the witness
yields a very good estimate.

Note that it is also possible to design a witness for the largest possible tolerance
to the noise in \EQ{nw}. Due to the special form of the noise, the fidelity estimate turns out to be equal to the fidelity.
This is remarkable: The fidelity can be
obtained exactly with only three local measurements.

\subsubsection{Measuring the projector-based witness}

\begin{figure}
\centerline{ \epsfxsize 5cm \epsffile{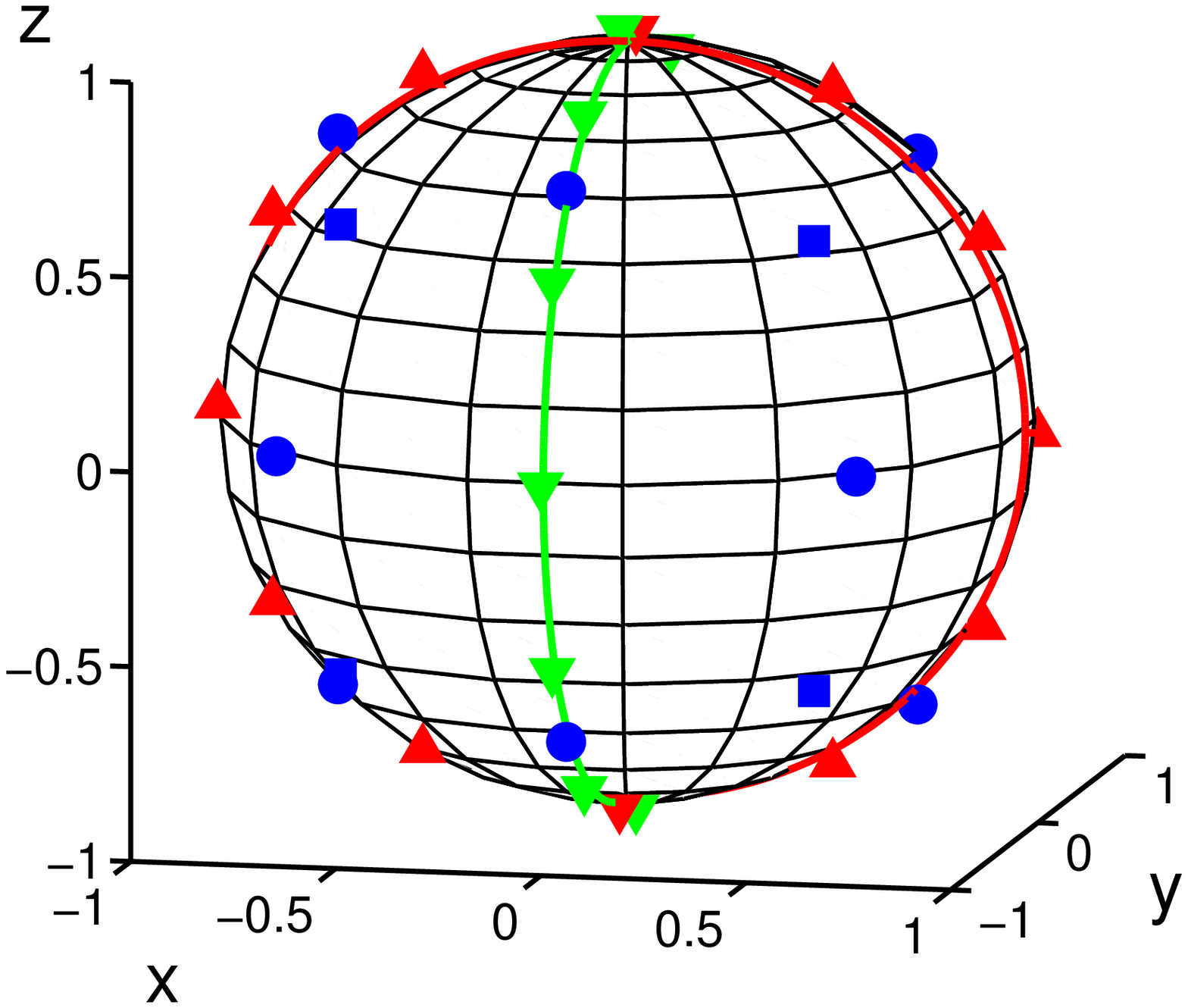} \epsfxsize 5cm \epsffile{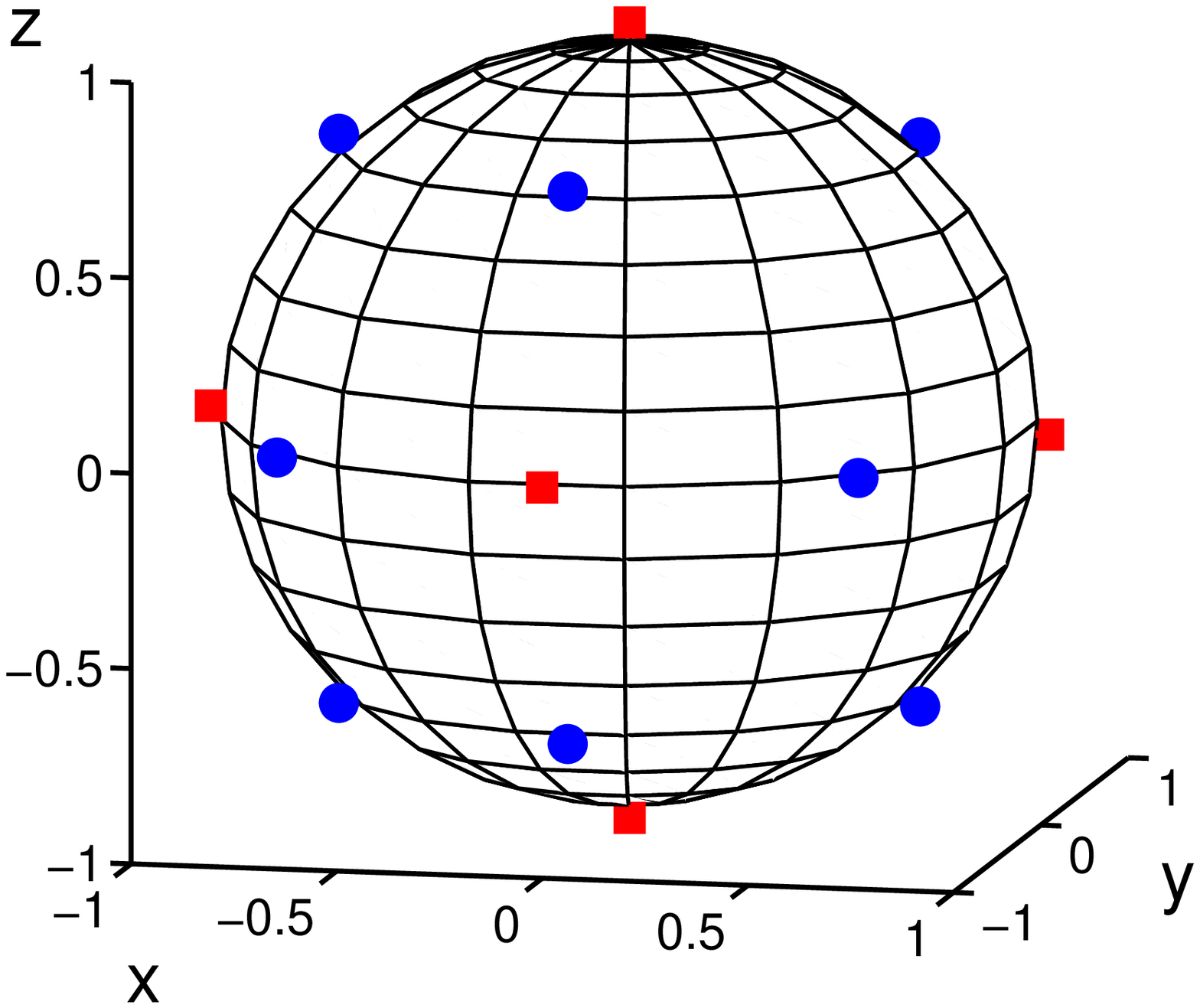}}
\centerline{\hskip 0.5cm (a) \hskip 4.5cm (b)}
\caption{(a) The measurement settings needed to measure the projector to the six-qubit symmetric Dicke state with three excitations based on the decomposition \EQ{decompDicke6}. A point at $(x,y,z)$ indicates measuring $x\sigma_x+y\sigma_y+z\sigma_z$ on all qubits. ($\bigtriangleup$) Settings for ${\rm Mermin}_{x,z}$, ($\bigtriangledown$) settings for ${\rm Mermin}_{y,z}$, ($\square$) $\sigma_x \pm \sigma_y \pm \sigma_z$ , and ($\bigcirc$) rest of the settings. (b) Settings for the four-qubit Dicke state with two excitations based on \EQ{decompDicke4}. ($\square$) $\pm\sigma_x, \pm \sigma_y, \pm \sigma_z$ , and ($\bigcirc$) $\sigma_x\pm\sigma_y, \sigma_x\pm\sigma_z,$ and $\sigma_y\pm\sigma_z.$} \label{fig_meas_settings}
\end{figure}

For measuring the projector-based witness \EQ{WPD} for $N=6,$
one has to decompose the projector in an efficient way.
The straightforward decomposition into the weighted sum of products of Pauli spin matrices
leads to a scheme that needs $183$ settings, since for
all local operators all the permutations have to be measured.
The number of settings needed can dramatically be decreased if
one is looking for a decomposition of the form \EQ{dec}. Observation 1 makes it possible
to decompose the projector in this way such that only 25 settings are needed.
We could further decrease the number of settings needed and
 found the following decomposition
\begin{eqnarray}
64\ketbra{D_6^{(3)}}&=&
-0.6[\openone]+0.3[x\pm\openone]-0.6[x]
+0.3[y\pm\openone]-0.6[y]\nonumber\\
&&+0.2[z\pm\openone]-0.2[z]+0.2 {\rm Mermin}_{0,z} \nonumber\\
&&+0.05[x\pm y\pm\openone]-0.05[x\pm z\pm\openone]-0.05[y\pm
z\pm\openone]
\nonumber\\&&
-0.05[x\pm y\pm z]+0.2[x\pm z]+0.2[y\pm z]+0.1[x\pm y]
\nonumber\\
&&+0.6{\rm Mermin}_{x,z}+0.6{\rm Mermin}_{y,z}.
\label{decompDicke6}
\end{eqnarray}
Here we use the notation
$[x+y]=(\sigma_x+\sigma_y)^{\otimes 6},$
$[x+y+\openone]=(\sigma_x+\sigma_y+\openone)^{\otimes 6},$ etc.
The $\pm$ sign denotes a summation over the two signs, i.e.,
$[x\pm y]=[x+y]+[x-y].$
The Mermin operators are defined as
\begin{equation}
{\rm Mermin}_{a,b}:=\sum_{k \text{ even}} (-1)^{k/2} \sum_k \mathcal{P}_k (
\otimes_{i=1}^{k} \sigma_a  \otimes_{i=k+1}^{N}\sigma_b),
\end{equation}
where $\sigma_0=\openone.$ That is,
it is the sum of terms with even number of $\sigma_a$'s and
$\sigma_b$'s, with the sign of the terms depending on the number of
$\sigma_a$'s. The expectation value of the operators
${\rm Mermin}_{a,b}$ can be measured based on the decomposition \cite{toolbox}
\begin{equation}
{\rm Mermin}_{a,b}=\frac{2^{N-1}}{N} \sum_{k=1}^N (-1)^k
\bigg[\cos\left(\tfrac{k\pi}{N}\right)a+\sin\left(\tfrac{k\pi}{N}\right)b\bigg]^{\otimes
N}.
\end{equation}
Hence, ${\rm Mermin}_{x,z}$ and ${\rm Mermin}_{y,z}$ can be measured
with six settings. ${\rm Mermin}_{0,z},$ on the other hand, needs only the measurement of the $\{\sigma_z,\sigma_z,\sigma_z,\sigma_z,\sigma_z,\sigma_z\}$ setting. Knowing that $[A]$, $[A+\openone]$ and
$[A-\openone]$ can be measured with a single setting $\{A,A,A,...,A\}$, we find that $21$ measurement settings are needed to measure $\ketbra{D_6^{(3)}}:$
$x, y, z, x\pm y, x\pm z, y\pm z, \sqrt{3}x\pm z, \sqrt{3}z\pm x, \sqrt{3}y\pm z, \sqrt{3}z\pm y,$ and
$ x \pm y \pm z.$ The settings are also shown in figure~\ref{fig_meas_settings}(a) \footnote{Note that \REF\cite{dicke6experiment2} presents another decomposition that needs also $21$ settings.}.

\subsection{Witness independent from the projector witness}

So far we constructed witnesses that detected fewer states than the projector-based witness, in return, they were easier to measure. When proving that they were witnesses, we used the simple relation \EQ{WWWP}. Following the example of \REF\cite{T07}, we now look for a  two-setting
witness of the form \EQ{WI2} for $N=6$ that is independent from the projector witness.
For determining $c_{D6},$
we need to compute the maximum of $J_x^2+J_y^2$ for biseparable states for all the possible bipartitions.
%Since $J_x^2+J_y^2$ is permutationally symmetric, it is sufficient to look for the three different %bipartitions $(1)(23456),(12)(3456),(123)(456).$
As we have discussed in section~\ref{sec_w_indep_pw}, instead of looking for the maximum for states that are separable with respect to a certain bipartition, we can also look for the maximum for PPT states.  (See Appendix A.)
We obtain
 \be
 c_{D6}:=11.0179.
 \ee
$\WW_{\rm D(6,3)}^{(I2)}$ detects genuine multipartite entanglement if for white noise $p_{\rm noise}< 0.1091.$ Simple numerical optimization leads to the same value for the maximum
for biseparable states %\cite{boptim}
\footnote{We used the \texttt{maxbisep}
routine of the QUBIT4MATLAB V3.0 package \cite{QUBIT4MATLAB} with parameters for accuracy
$[30000, 100000,0.0005].$ See also Appendix B.}. Hence we find that
our witness is optimal. Finally, the list of witnesses presented in this section are shown in the top part of table 1.

\begin{table}[htdp]
\caption{The list of entanglement witnesses presented in this paper, together with the number of measurement settings needed to measure them and their robustness to white noise. Top four lines: six-qubit witnesses. Bottom five lines: four- and five-qubit witnesses.}
\begin{center}
\begin{tabular}{ccc}
%\hline
\br
Witness & Number of settings & Noise tolerance \\
\hline
$\WW_{\rm D(6,3)}^{(P)}$ & $21$ & $0.4063$ \\
$\WW_{\rm D(6,3)}^{(P3)}$ & $3$ & $0.2735$ \\
$\WW_{\rm D(6,3)}^{(P2)}$ & $2$ & $0.1391$ \\
$\WW_{\rm D(6,3)}^{(I2)}$ & $2$ & $0.1091$ \\
% \hline
$\WW_{\rm D(5,2)}^{(I2)}$ & $2$ & $0.1046$ \\
$\WW_{\rm D(4,1)}^{(P)}$ & $7$ & $0.2667$ \\
$\WW_{\rm D(4,1)}^{(I2)}(q=1.47)$ & $3$ & $0.1476$ \\
$\WW_{\rm D(4,2)}^{(P)}$ & $9$ & $0.3556$\\
$\WW_{\rm D(4,2)}^{(P3)}$ & $2$ & $0.2759$ \\
%\hline
\br
\end{tabular}
\end{center}
\label{wittab}
\end{table}%

\section{Witnesses for states derived from $\ket{D_6^{(3)}}$ via projections}

By projective measurements of one or two of the qubits
we can obtain several states that are inequivalent
under stochastic local operations and classical communication (SLOCC).
Surprisingly, these states still possess genuine multipartite entanglement \cite{dicke6experiment,WK09}.
Next, we discuss how to detect the entanglement of these states.

\subsection{Witnesses for the superposition of five-qubit Dicke states: }

After measuring one of the qubits in some basis and post-selecting for one of the two outcomes,
one can obtain states of the form
 \be
  \varrho_{\rm D5}:=c_1\ket{D_5^{(2)}}+c_2\ket{D_5^{(3)}}, \label{D5}
  \ee
 where $\vert c_1\vert^2+\vert c_2\vert^2=1.$
 For such states, the expectation value of $J_x^2+J_y^2$ is maximal,
 thus a witness of the form \EQ{WI2} for $N=5$
 is used to detect their entanglement.
Both semidefinite programing and simple numerical optimization leads to
$
c_{D5}:=7.8723.
$
Naturally,  $\exs{\WW_{\rm D5}^{(I2)}}$ is minimal not only for states of the form \EQ{D5}, but for any mixture of such states.

\subsection{Witness for the four-qubit W-state}

\begin{figure}
\centerline{ \epsfxsize 9cm \epsffile{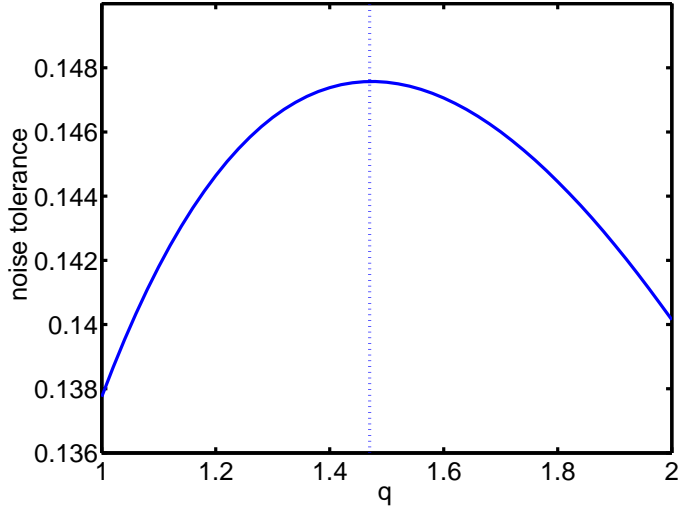}}
\caption{The noise tolerance of the witness  $\WW_{D(1,4)}^{(I3)}$ given in \EQ{WI3} as a function of the parameter $q.$ The maximum is in the vicinity of $q=1.47.$} \label{fig_w4}
\end{figure}

Now we will construct witnesses for a four-qubit W-state, which is obtained from $\ket{D_6^{(3)}}$ if two qubits are measured in the $\sigma_z$ basis, and the measurement result is $+1$ in both cases.
We consider a witness of the form \EQ{WI3} for $N=4$ and $m=1.$
 We try several values for $q$ and determine $c_q$ for the witness
 $\WW_{\rm D(4,1)}(q)$ as a function of $q$
 using semidefinite programming. For each witness we also compute the noise tolerance.
 The results of these computations can be seen in figure~\ref{fig_w4}. It turns out, that the best witness is obtained for $q=1.47$ and $c_q=4.1234.$ It tolerates white noise if $p_{\rm noise}<0.1476.$

\subsection{Three-setting witness for the four-qubit Dicke state}

A $\ket{D_4^{(2)}}$ state can also be obtained from $\ket{D_6^{(3)}}$,
namely if the measurement outcomes are $+1$ and $-1$ for two consecutive $\sigma_z$ measurements.
For that case,
we look for a three-setting  witness, based on the projector witness.
For white noise, the result  is
 \be
 \WW_{\rm D(4,2)}^{(P3)}:=2\cdot\openone +
 \tfrac{1}{6}(J_x^2+J_y^2-J_x^4-J_y^4)+\tfrac{31}{12}J_z^2-\tfrac{7}{12}J_z^4.
 \ee
The witness tolerates white noise if $p_{\rm noise}<0.2759.$
It is easy to check that $\WW$ is a witness: One has to notice that
$\WW_{\rm D(4,2)}^{(P3)}-3\WW^{(P)}_{\rm D(4,2)}\ge 0,$ were $\WW^{(P)}_{\rm D(4,2)}$ is defined in
\EQ{WPD}. Thus, the fidelity can be estimated from the measurement of the witness as
$F\ge 2/9-\exs{\WW_{\rm D(4,2)}^{(P3)}}/3.$

\subsection{Measuring the projector witness for the four-qubit Dicke state}

We can also measure the projector witness $\WW_{\rm D(4,2)}^{(P)}=\tfrac{2}{3}\openone-\ketbra{D_4^{(2)}}.$ The method in Observation 1 gives the following decomposition for the projector
\begin{eqnarray}
16\ketbra{D_4^{(2)}}
&=&\tfrac{2}{3}([x]+[x\pm\openone]+[y]+[y\pm\openone])\nonumber\\
&+&\tfrac{1}{3}(8[z]-[z\pm\openone]-[x\pm z]-[y\pm z])
+\tfrac{1}{6}[x\pm y].\label{decompDicke4}
\end{eqnarray}
The $9$ measurement settings are
$x,y,z, x\pm y,x\pm z,$ and $y\pm z,$  shown also in figure~\ref{fig_meas_settings}(b). The list of witnesses presented in this section are given in the bottom part of table 1.

\section{Conclusions}

In summary, we presented general methods for constructing entanglement witnesses for detecting genuine multipartite entanglement in experiments. In particular, we considered projector-based witnesses and found efficient decompositions for them. Then, we constructed two- and three-setting
witnesses for symmetric Dicke states
that were based on the projector witness, as well as independent from the
projector witness. We applied our methods to design witnesses for the recent experiment observing a six-qubit symmetric Dicke state with three excitations \cite{dicke6experiment}. Our methods can be generalized for future experiments. As a first step, in Appendix C we list some entanglement witnesses for systems with $5-10$ qubits. Moreover, recent results on the symmetric tensor rank problem suggest that decompositions more efficient than the one in Observation 1 are possible, however, they involve complex algorithms \cite{tensor_rank}. Thus, it would be interesting to look for better upper bounds for the number
of settings used for symmetric operators.

% \section*{Acknowledgments}

\ack

We thank A. Doherty, O. G\"uhne, D. Hayes, and S. Pironio for fruitful discussions. We
thank the support of the DFG-Cluster of Excellence MAP, and the EU (QAP, SCALA).
W.W. acknowledges the support by QCCC of the Elite Network of Bavaria and
the Studientiftung des dt. Volkes. G.T. thanks the National
Research Fund of Hungary OTKA (Contract No. T049234),
the Hungarian Academy of Sciences (Bolyai Programme),
and the Spanish MEC (Ramon y Cajal Programme,
Consolider-Ingenio 2010 project ''QOIT'').

\appendix

\section{Semidefinite programming used for obtaining witnesses}

Here we summarize two optimization problems that are useful for designing entanglement witnesses
and can be solved with semidefinite programming. Both tasks are related to
designing witnesses that are easy to measure.

1. Semi-definite programming can be used to find the witness $\WW$ with
the largest noise tolerance as explained in the beginning of section~\ref{sec_ts_bp}.
The corresponding task can be formulated as
\begin{equation}
  \begin{array}{ll}
    \text{{\bf minimize}} & \sum_kc_k\trace(B_k\varrho_{\rm noise}), \\
    \\
    \text{{\bf subject to}} & \sum_kc_k\trace(B_k\varrho)=-1, \\
                            & \sum_kc_kB_k-\alpha \WW^{(P)}\ge 0, \\
                            & \alpha>0. \\
  \end{array}
  \label{taskoptim}
\end{equation}
Here $\varrho$ is the state around which we detect entanglement.
$\varrho_{\rm noise}$ is the noise, not necessarily white. The
optimization is over $\alpha$ and the $c_k$'s.

2. Semi-definite programming can be used to look for the maximum for $PPT$ bipartite states. This gives an upper bound on the maximum for biseparable states. In many cases, the two   coincide.  The corresponding task can be formulated as a
standard semidefinite
program as
\begin{equation}
  \begin{array}{ll}
    \text{{\bf minimize}} & -\trace(M\varrho), \\
    \\
    \text{{\bf subject to}} & \varrho\ge 0, \\
                            & \trace(\varrho)=1,\\
                            & \varrho^{T_A}\ge 0. \\
  \end{array}
\end{equation}
Here $T_A$ means partial transpose according to some groups of the qubits.

\section{List of MATLAB subroutines}

We summarize some of the MATLAB routines of the QUBIT4MATLAB 3.0 package that
can be used for the calculations necessary for designing entanglement witnesses. A full list of the commands is given in \REF\cite{QUBIT4MATLAB}.

The command \texttt{decompose} can be used to obtain a decomposition of a Hermitian operator into the sum of products of Pauli spin matrices.
Moreover, \texttt{maxsep} and \texttt{maxsymsep} can be used for getting the maximum for separable multi-qudit states and symmetric product states for a Hermitian operator, respectively. The command \texttt{maxbisep} gives the maximum for states that are biseparable with respect to some partitioning of the qubits. The command \texttt{maxb} gives the maximum for all possible bipartitions. All these commands look for the maximum with a simple optimisation algorithm that is not guaranteed to find the global maximum, nevertheless, it typically does find it. \texttt{overlapb} gives the maximum overlap of a state $\ket{\Psi}$ and biseparable states. It can be used to construct entanglement witnesses of the type \EQ{overlap}.

For semidefinite programming, we used SeDuMi \cite{sedumi} and YALMIP \cite{yalmip}.
Two subroutines based on them are now in QUBIT4MATLAB 4.0 \cite{QUBIT4MATLAB40}.
The command
 \texttt{optwitness} looks for the best witness that can be composed linearly from a set of operators, while
 \texttt{maxppt} determines the maximum of an operator expectation value for states with a positive partial transpose for some bipartitioning of the qubits.

\section{Witnesses for systems with $5-10$ qubits}

A three-setting witness based on the projector witness for the state $\ket{D_8^{(4)}}$ is given by
\be
 \WW_{D(8,4)}^{(P3)}:=1.3652\cdot\openone + \sum_{l=x,y,z} \sum_{n=1}^4
 c_{ln} J_l^{2n}, \ee with \be \{c_{ln}\}=\left(
 \begin{array}{cccc}
         0.0038612 & -0.0052555 &  0.0015016 & -0.00010726\\
          0.0038612  &-0.0052555&   0.0015016  &-0.000107266\\
          3.124  &-1.07699&   0.11916  &-0.0038992\\
 \end{array}
  \right).
 \ee
The noise-tolerance for white noise is $p_{\rm noise}<0.2578.$  For larger $N,$ we can use the ansatz
\be
\fl \WW_{D(N,N/2)}^{(P3)}:=c_1\cdot\openone + c_{xy}
  \{(\sigma_x+\openone)^{\otimes N}+(\sigma_x-\openone)^{\otimes N}+(\sigma_y+\openone)^{\otimes N}+(\sigma_y-\openone)^{\otimes N}\}
  +\sum_{n=1}^{N/2} c_{zn} J_z^{2n}.
\ee
For $N=10,$ the optimal coefficients are $c_1=1.3115, c_{xy}=-0.0023069,$
and $c_z=\{3.4681,-1.2624,0.16494,-0.0084574,0.000146551\}.$
White noise is tolerated if $p_{\rm noise}<0.2404.$
The large noise-tolerance for the $N=10$ case suggests that a robust three-setting witness for $\ket{D_{N}^{(N/2)}}$ might be constructed even for large $N.$

A three-setting witness independent of the projector witness for the $N$-qubit $W$-state is given by
\EQ{WI3} for $m=1.$
For $N=5,$ we have $c_5=5.6242$, $q_5=2.22,$ and the witness tolerates white noise if $p_{\rm noise}<0.0744.$
For $N=6,$ we have $c_6=7.1095$, $q_6=3.13,$ and noise is tolerated if $p_{\rm noise}<0.0401.$


\section*{References}

\begin{thebibliography}{99}

\bibitem{reviews} For reviews see
Horodecki R, Horodecki P, Horodecki M, and Horodecki K
2009 \emph{Rev. Mod. Phys.} \textbf{81} 865
Plenio M and Virmani S
2007 \emph{Quant. Inf. Comp.} \textbf{7} 1.

\bibitem{GT08} G\"uhne O and  T\'oth G 2009 \emph{Phys. Rep.} {\bf 474} 1.

%%%%%%% QI experiments

% 3-qubit GHZ state with photons
\bibitem{PB00}
Pan J W, Bouwmeester D, Daniell M, Weinfurter H, and Zeilinger A
2000 \emph{Nature} {\bf 403} 515.

\bibitem{BE04} Bourennane M, Eibl M, Kurtsiefer C, Gaertner S, Weinfurter H, G\"uhne O, Hyllus Ph,
Bru{\ss} D, Lewenstein M, and Sanpera A
    2004 \emph{Phys. Rev. Lett.} {\bf 92} 087902.

\bibitem{KS05} Kiesel N, Schmid C, Weber U, T\'oth G, G\"uhne O, Ursin R, and Weinfurter H
2005 \emph{Phys. Rev. Lett.} \textbf{95} 210502.

% Dicke state
\bibitem{KS07} Kiesel N, Schmid C, T\'oth G, Solano E, and Weinfurter H
 2007 \emph{Phys. Rev. Lett.} {\bf 98} 063604.

% Four-photon state
\bibitem{WS08} Wieczorek W, Schmid C, Kiesel N,
Pohlner R, G\"uhne O, and Weinfurter H
2008 \emph{Phys. Rev. Lett.} {\bf 101} 010503.

% Six photons
\bibitem{LZ07} Lu C Y, Zhou X Q, G\"uhne O, Gao W B,
Zhang J, Yuan Z S, Goebel A, Yang T, and Pan J W
2007 \emph{Nat. Phys.} {\bf  3} 91.

% Ten qubit photonic
\bibitem{GL08} Gao W B, Lu C Y, Yao X C, Xu P, G\"uhne O,
Goebel A, Chen Y A, Peng C Z, Chen Z B, and Pan J W
%Experimental demonstration of a hyper-entangled ten-qubit Schrodinger cat state (2008),
2008 \emph{Preprint} arXiv:0809.4277.

%%%%%%%%%%%%

% 4-qubit GHZ state with ions
\bibitem{SK00} Sackett C A, Kielpinski D, King B E, Langer C, Meyer V,
Myatt C J,  Rowe M, Turchette Q A, Itano W M,  Wineland D J, and
Monroe C
2000 \emph{Nature} {\bf 404} 256.

% Six ions
\bibitem{LK05} Leibfried D, Knill E, Seidelin S, Britton J, Blakestad R B, Chiaverini J, Hume D B,
Itano W M, Jost J D, Langer C, Ozeri R, Reichle R, and Wineland D J
    2005 \emph{Nature} {\bf 438} 639.

% Eight ions
\bibitem{HH07} H\"affner H, H\"ansel W, Roos C F, Benhelm J, Chek-al-kar D, Chwalla M, K\"orber T, Rapol U D, Riebe M, Schmidt P O,
    Becher C, G\"uhne O, D\"ur W, and Blatt R
    2005 \emph{Nature} {\bf 438} 643.

% Cold atoms
\bibitem{KB98} Hald J, S\o rensen J L, Schori C, and Polzik E S
1999 \emph{Phys. Rev. Lett.} {\bf 83} 1319.

% Cluster state in an optical lattices
\bibitem{MG03} Mandel O, Greiner M, Widera A, Rom T, H\"ansch T, and Bloch I
2003 \emph{Nature} {\bf 425} 937.

% diamond, nuclear and electronic spins are entangled
\bibitem{NM08} Neumann P, Mizuochi N, Rempp F, Hemmer P, Watanabe H, Yamasaki S, Jacques V, Gaebel T, Jelezko F, and Wrachtrup J
2008 \emph{Science} {\bf  320} 1326.

%%%%%%%%%%%%%%%%%%%

\bibitem{T02} Terhal B M
2002 \emph{Theoret. Comput. Sci.} \textbf{287} 313.

% Optimization of entanglement witnesses
\bibitem{LK00} Lewenstein M,
Kraus B, Cirac J I, and Horodecki P
2000 \emph{Phys. Rev. A} {\bf 62} 052310.

\bibitem{GH02} G\"uhne O, Hyllus P, Bru{\ss} D, Ekert A, Lewenstein M,
Macchiavello C, and Sanpera A
2002 \emph{Phys. Rev. A} \textbf{66} 062305.

\bibitem{GH03} G\"uhne O, Hyllus P, Bruss D, Ekert A, Lewenstein M, Macchiavello C, and
Sanpera A
2003 \emph{J. Mod. Opt.} {\bf 50} 1079.

%%%%%%%%%%%%%%%%%%%

% Bells theorem without inequalities
\bibitem{GH90} Greenberger D M, Horne M A, Shimony A, and Zeilinger A
1990 \emph{Am. J. Phys.} {\bf 58} 1131.

%ÒPersistent entanglement
%in arrays of interacting particlesÓ
\bibitem{BR01} Briegel H J and Raussendorf R
2001 \emph{Phys. Rev. Lett.} {\bf 86} 910.

%%%%%%%%%%%%%%%%%

\bibitem{TG05} T\'oth G and G\"uhne O
2005 \emph{Phys. Rev. Lett.} {\bf 94}, 060501 (2005).

\bibitem{toolbox} G\"uhne O, Lu C Y, Gao W B, and Pan J W
2007  \emph{Phys. Rev. A} {\bf 76}, 030305 (2007).

%%%%%%%%%%%%%%%%%%%

\bibitem{DP02} Doherty A C, Parrilo P A, and Spedalieri F M
2002 \emph{Phys. Rev. Lett.}  {\bf 88} 187904;
2004 \emph{Phys. Rev. A} 69, 022308 .

\bibitem{BV04} Brand\~ao F G S L and Vianna R O
%"Separable Multipartite Mixed States: Operational Asymptotically
%Necessary and Sufficient Conditions",
2004 \emph{Phys. Rev. Lett.} {\bf 93} 220503.

\bibitem{EH04} Eisert J, Hyllus P, G\"uhne O, and Curty M
%"Complete hierarchies of efficient approximations to problems in
%entanglement theory",
2004 \emph{Phys. Rev. A} {\bf 70} 062317.

%Manipulating multiqudit entanglement witnesses by using linear programming
\bibitem{JN07} Jafarizadeh M A, Najarbashi G, and Habibian H
2007 \emph{Phys. Rev. A} {\bf 75} 052326 (2007).

\bibitem{HE08} Hyllus P and Eisert J
2006 \emph{New J. Phys.} {\bf 8} 51 (2006).

\bibitem{WP09} Wunderlich H and Plenio M B
2009 \emph{Preprint} arXiv:0902.2093.

%%%%%%%%%%%%

\bibitem{Dicke} Dicke R H 1954 \emph{Phys. Rev.} {\bf 93} 99.

%"Characterizing the entanglement of symmetric many-particle spin-1/2
%systems"
\bibitem{SG03} Stockton J K, Geremia J M, Doherty A C, and Mabuchi H
2003 \emph{Phys. Rev. A} {\bf 67} 022112.

\bibitem{dicke6experiment} Wieczorek W, Krischek R, Kiesel N,
Michelberger P, T\'oth G, and Weinfurter H 2009 \emph{Phys. Rev. Lett.} \textbf{103} 020504.

\bibitem{dicke6experiment2} For another experiment aiming to observe a six-qubit Dicke state see
Prevedel R, Cronenberg G, Tame M S, Paternostro M, Walther P, Kim M S, and Zeilinger A
2009 \emph{Phys. Rev. Lett.} \textbf{103} 020503. See also the related theoretical work Campbell S, Tame M S, Paternostro M  2009 \emph{New J. Phys.} \textbf{11}  073039.

\bibitem{QUBIT4MATLAB} T\'oth G
2008 \emph{Comput. Phys. Comm.} {\bf 179} 430.

\bibitem{AB01} Ac\'{\i}n A, Bru{\ss} D, Lewenstein M, and Sanpera A 2001 \emph{Phys. Rev. Lett.} {\bf 87} 040401.

\bibitem{phdguhne} G\"uhne O
2004 \emph{Ph.D. Thesis}, University of Hannover.

%\bibitem{decomp} Other decompositions are also possible. In particular, instead of binary $s_n$ variables
%one can use $s_n\in\{\exp(i2\pi k/S)\}_{k=1}^S.$ For the $S=N$ case, one can write
%$A=\sum_n c_nN^{-(N-1)}[\sum_{s_1s_2s_3...=1} (s_1B_1+s_2B_2+...)^{\otimes N}-\sum_k B_k^{\otimes N}].$ A %decomposition with continous number of terms is of the form
%$\sum_k P_k (B_1 \otimes B_2 \otimes ...)P_k \propto \int_{ %\phi_k\in[0,2\pi]}[e^{i\phi_1}B_1+e^{i\phi_2}B_2+...+e^{i\phi_{N-1}}B_{N-1}+e^{-i(\phi_1+\phi_2+...+\phi_{N-1})}B_N]^{\otimes %N}d\phi_1 d\phi_2...d\phi_{N-1}.$ The coefficients are complex thus one has to use that
%$[(A+iB)^{\otimes N}+(A-iB)^{\otimes N}]/2$ with Hermitian $A$ and $B$ is the Mermin operator, which can be %decomposed
%into the sum of $N$ local terms \cite{toolbox}.

\bibitem{T07} T\'oth G 2007 \emph{J. Opt. Soc. Am. B} {\bf 24} 275.

\bibitem{ppt}
Peres A,
1996 \emph{Phys. Rev. Lett} {\bf 77} 1413;
Horodecki M, Horodecki P, and Horodecki R
1996 \emph{Phys. Lett. A} {\bf 223} 1.

%%%%%%%%%%%%%%%%%

\bibitem{structure} Witnesses for the state $\ket{D_6^{(3)}}$
are presented with the structure factor
 in Krammer P, Kampermann H,  Bruss D,
Bertlmann R A, Kwek L C, and Macchiavello C
2009 \emph{Preprint} arXiv:0904.3860.
In a sense, these witnesses are written with collective quantities, after a site-dependent phase shift is applied.

%%%%%%%%%%%%%%%%%

\bibitem{UF03} Unanyan R G and Fleischhauer M
%"Decoherence-Free Generation of Many-Particle Entanglement by Adiabatic Ground-State
%Transitions",
2003 \emph{Phys. Rev. Lett.} {\bf 90} 133601.

\bibitem{DK03} Duan L M and Kimble H J
%"Efficient Engineering of Multiatom Entanglement through
%Single-Photon Detections",
2003 \emph{Phys. Rev. Lett.} {\bf 90} 253601.

\bibitem{TZ07} Thiel C, von Zanthier J, Bastin T, Solano E, and Agarwal G S
%``Generation of Symmetric Dicke States of Remote Qubits with Linear Optics'',
2007 \emph{Phys. Rev. Lett.} {\bf 99},  193602.

% Deterministic Dicke state preparation with continuous measurement and control
\bibitem{SH04} Stockton J K, van Handel R, and Mabuchi H
2004 \emph{Phys. Rev. A} {\bf 70} 022106.

%%%%%%%%%%%%%%%%

%\bibitem{alternative_form} The witnesses can also be given in an alternative form \EQ{permform} as
%$\WW_{D(6,3)}^{(P2)}:=396\cdot\openone-10(M_{2x}+M_{4x}+M_{2y}+M_{4y})-50(M_{6x}+M_{6y})$
%and
%$\WW_{D(6,3)}^{(P3)}:=170\cdot\openone-4(M_{2x}+M_{4x}+M_{6x}+M_{2y}+M_{4y}+M_{6y})+6(M_{2z}-M_{4z})+46M_{6z},$
%where $M_{kl}=\sum_n \mathcal{P}_n [\sigma_l^{\otimes k}\otimes \openone^{\otimes (N-k)}].$

%%%%%%%%%%%%%%%%

%\bibitem{boptim} We use the \texttt{maxbisep}
%routine of the QUBIT4MATLAB V3.0 package \cite{QUBIT4MATLAB} with parameters for accuracy
%$[30000, 100000,0.0005].$ See also Appendix B.

\bibitem{WK09} Wieczorek W, Kiesel N, Schmid C, and Weinfurter H 2009 \emph{Phys. Rev. A} {\bf 79} 022311.

\bibitem{tensor_rank} Comon P, Golub G, Lim L H, and Mourrain B
2008 \emph{SIAM Journal on Matrix Analysis Appl.} {\bf 30} 1254.

\bibitem{sedumi} Sturm J, SeDuMi, a MATLAB toolbox for optimization over symmetric cones, URL
http://sedumi.mcmaster.ca.

\bibitem{yalmip} L\"ofberg J, Yalmip: A toolbox for modeling and optimization in MATLAB,\\
URL http://control.ee.ethz.ch/$\sim$joloef/yalmip.php.

\bibitem{QUBIT4MATLAB40} http://www.mathworks.com/matlabcentral/fileexchange/8433.

\end{thebibliography}
\end{document}